\documentclass[12pt,preprint]{aastex}   
\newcommand{\Mmi}{\hbox{~\rm{$\mbox{Mm}^{-1}$}}}
\newcommand{\muHz}{\hbox{$~\rm{\mu Hz}$}}
\newcommand{\mwd}{\hbox{\rm{$\Gamma$}}}
\newcommand{\Ux}{\hbox{\rm{$U_x$}}}
\newcommand{\Uy}{\hbox{\rm{$U_y$}}}
\newcommand{\ux}{\hbox{\rm{$u_x$}}}
\newcommand{\uy}{\hbox{\rm{$u_y$}}}

\shorttitle{VARIATIONS IN $P$-MODE PARAMETERS WITH A LARGE FLARE}
\shortauthors{MAURYA, AMBASTHA AND TRIPATHY}

\begin{document}

\title{VARIATIONS IN $p$-MODE PARAMETERS WITH CHANGING ONSET-TIME OF A LARGE FLARE}

\author{R. A. MAURYA and A. AMBASTHA}
\affil{Udaipur Solar Observatory, Physical Research Laboratory, Udaipur-313001, INDIA.}

\and

\author{
S. C. TRIPATHY}
\affil{National Solar Observatory, 950 N Cherry Avenue, Tucson, AZ 85719, USA}

\email{ramajor@prl.res.in, ambastha@prl.res.in, stripathy@noao.edu}


\begin{abstract}

It is expected that energetic solar flares releasing large amount of energy at the photosphere may be able to excite the acoustic ($p$-) modes of oscillations. We have determined the characteristic properties of mode parameters by applying the ring diagram technique to 3-D power spectra obtained for solar active region NOAA 10486 during the long duration energetic X17.2/4B flare of October 28, 2003. Strong evidence of substantial increase in mode amplitude and systematic variations in sub-surface flows, i.e., meridional and zonal components of velocity, kinetic helicity, vorticity, is found from comparison of the pre- to the post-flare phases.

\end{abstract}

%

\section{Introduction}
\label{sec:introduction}

\citet{1962ApJ...135..474L} discovered solar 5-minute oscillations while measuring Doppler shifts in photospheric spectral lines. \citet{1970ApJ...162..993U} and \citet{1971ApL.....7..191L} later interpreted these oscillations to be due to acoustic (or $p$-) modes trapped in the sub-photospheric layers. Accurate determination of these modes provides a powerful diagnostic tool for probing the solar interior. It is generally believed that these modes may be either intrinsically overstable or stochastically excited by turbulent convection. 

Characteristics of the surface $p$-modes and sub-surface flows are essentially described by shorter wavelength modes that are trapped near the surface. These modes can be studied using the ring diagram analysis which uses three dimensional data cubes (16\arcdeg$\times$16\arcdeg$\times$1664), first two correspond to spatial size of the active region and the third is time in minutes \citep{1988ApJ...333..996H}. Other techniques of local helioseismology, generally used for studying various aspects of sub-photospheric characteristics of active region interiors, are time-distance analysis \citep{2004ApJ...603..776Z}, and seismic holography \citep{2004ESASP.559..337B}.

\cite{1972ApJ...176..833W} first suggested that energetic flares may be able to excite acoustic waves by exerting mechanical impulse of the thermal expansion of the flare on the photosphere. They estimated the damping times to be longer than a day for the free modes. Subsequently, active region (AR) related effects on $p$-mode parameters have been reported by several researchers \citep{1990Natur.345..779L, 2000MNRAS.313...32C, 2000SoPh..192..363H, 2001ApJ...563..410R, 2003SoPh..218..151A,2004ESASP.559..293A, 2004ApJ...608..562H}. \cite{1998Natur.393..317K, 1999SoPh..190..459K} and \cite{1999ApJ...513L.143D} have also reported excitation of flare-related waves on the solar surface, however, these pertain to traveling waves as opposed to standing waves which constitute the normal modes of solar oscillations. 

The spatial extent of flares is usually much smaller than the spatial sizes of 16\arcdeg$\times$16\arcdeg~used in GONG and SOHO/MDI data-cubes for ring diagram analysis. Also, any mode amplification induced by transient flare events has to essentially compete with the absorption effects associated with intense magnetic fields of sunspots. Therefore it is expected to be rather difficult to conclusively resolve flare-related effects by averaging techniques. Here one can estimate the damping time scale of $p$-modes from the observed mode width, generally in the range of 15–-100\muHz~for high-degree modes. This gives a lifetime in the range of 0.5–-3 hr assuming that the width is mainly due to the finite life of the modes. Acoustic waves travel outwards from the flare site with the speed of sound varying in the range from 10 to 50 $\rm km~s^{-1}$ for the depths in which the high-degree modes are trapped. Thus, over the damping time of the order of an hour, these waves would travel a distance from 36 to 180 Mm, i.e., comparable to the spatial extent of the active region. A large fraction of energy dissipation occurs within this region if the flare is temporally located well within the data-cube. Therefore, it is expected that flare-related effects would last over several hours and should be detectable in the temporal averages carried out over a day as in the ring analysis. On the other hand, if the flare occurred around the end time of the data-cube, the flare effects would not be detectable as its effects would be contained mostly in the subsequent data-cube. 

Relation between photospheric motions and flare activity has been studied earlier. For example, correlation tracking studies have discovered a few cases of small-scale vorticity at the surface preceding a flaring event with timescales around an hour \citep{2004ApJ...617L.151Y}. Some cases of sunspot rotation in flaring active regions have also been found \citep{2003SoPh..216...79B}. Relationship of flare activity with some statistical properties of active regions has been studied by \cite{1995ApJ...446L.109L} and \cite{2005ApJ...619.1160A}, among others. It is believed that twisted flows in the sub-surface layers cause foot-point motions of sunspots observed at the photospheric layer leading to unstable magnetic topologies in the overlying regions. 

Although the average properties of $p$-mode parameters and sub-surface flows  have been studied with magnetic/flare activities by many researchers \citep{2004ESASP.559..293A, 2004ApJ...608..562H, 2005ApJ...631..636K, 2005ApJ...630.1184K, 2006ApJ...645.1543M}, the change of $p$-mode parameters and sub-surface flows during flares is not well understood. Some efforts have been made to examine the difference between flare productive and quiet regions to infer any activity related effects \citep{2006ApJ...645.1543M}. \cite{2003SoPh..218..151A,2004ESASP.559..293A} have found that power in $p$-modes appears to be larger during the period of high flare activity as compared to that in non-flaring regions of similar magnetic field strength. However, the pre-, peak and post-flare signatures of flares on $p$-mode parameters and sub-surface flows still require careful analysis. We have studied the long duration, energetic X17.2/4B flare of 28 October 2003 in NOAA 10486. We found strong evidence of flare-related changes in $p$-mode parameters and sub-surface flows by placing the flare at different temporal positions in appropriately constructed data-cubes.   

%
\section{The White Light Flare of NOAA 10486}
\label{S-ar10486}

\begin{figure}[ht] 
	\centering
		\includegraphics[width=0.95\textwidth,clip=,bb=20 11 531 288]{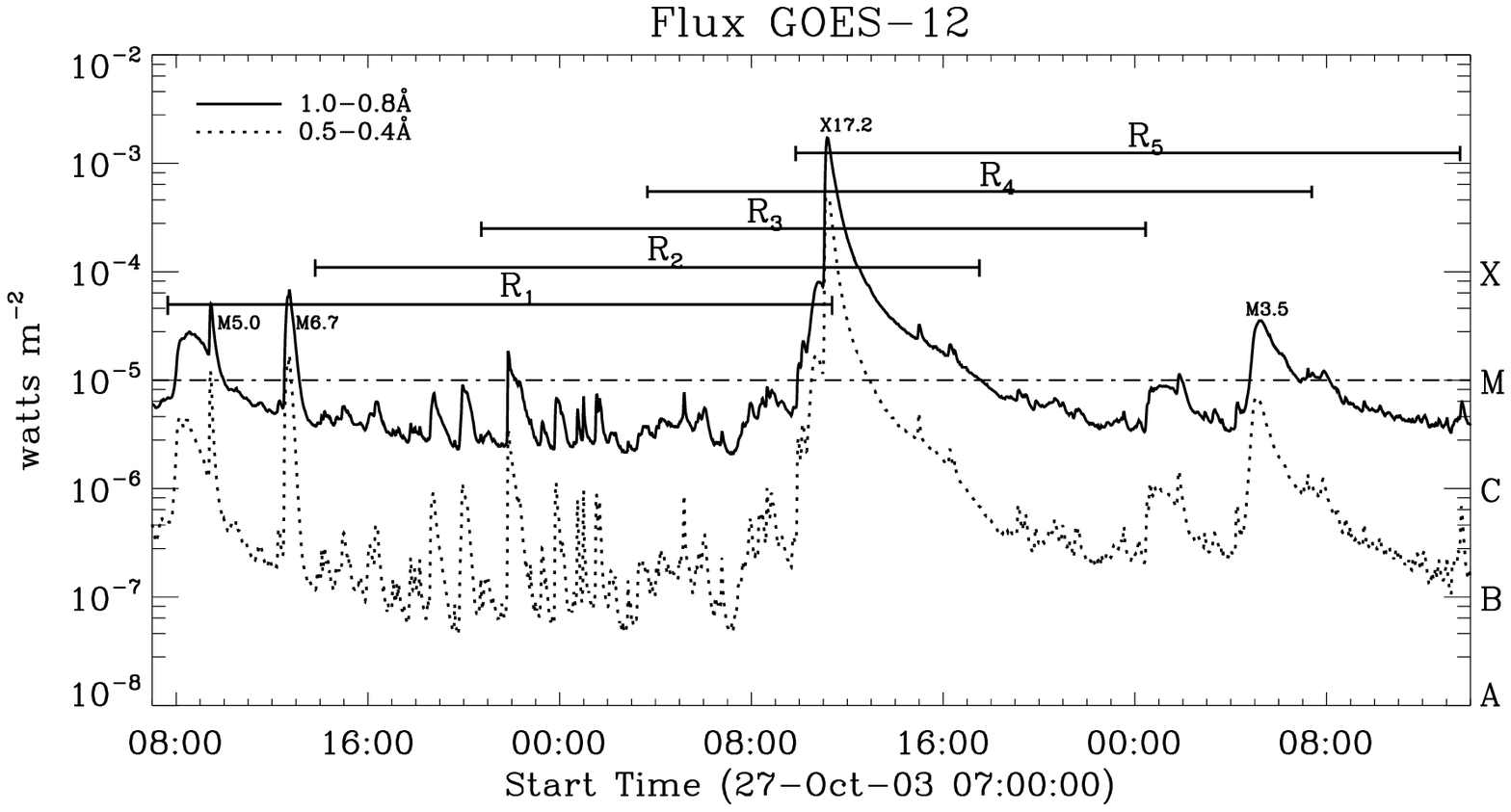}
	\caption{GOES-12 integrated X-ray flux for 1.0 - 0.8\AA~(solid line) and 0.5 - 0.4\AA~(dotted line). The horizontal lines with labels $R_i,  i=1, \ldots5,$ represent different time periods taken for constructing the five data-sets.}
	\label{fig:flare_datim}
\end{figure}

In order to determine any flare-related effects on $p$-mode parameters, we have considered one of the most energetic events that occurred in NOAA 10486, i.e., the white light super-flare of 28 October 2003, classified as X17.2/4B event. During the period of the data-sets used in our study, some other C and M class flares also occurred in NOAA 10486, but the X17.2/4B super flare was the dominant event during the 24-hr period.  According to Geostationary Operations Environmental Satellite (GOES) X-ray observations, this flare started at \mbox{09:51 UT}, reached the maximum phase at 11:10 UT, and decayed  at 11:24 UT, i.e., lasted over more than 90 minutes (Figure~\ref{fig:flare_datim}). However, it is evident that the integrated X-ray flux remained at a very high level well beyond 11:24 UT. Even if a background corresponding to the M1 level ($10^{-5}$ watts m$^{-2}$) is considered, the X-ray flux gradually reduced to this level only around 16:00 UT. In  H$\alpha$ also, it is reported to have lasted for more than four hours. Therefore, this long duration event (LDE) was a particularly well suited case for an investigation of flare-related helioseismic effects.
%
%
%
\section{The Data and Analysis} 
\label{S-data}
We have used high resolution GONG Dopplergrams obtained at one minute cadence. The  data-cubes have 1664 minutes' duration and cover 16\arcdeg$\times$16\arcdeg~area centered at 285\arcdeg~Carrington longitude and -22.5\arcdeg~latitude covering NOAA 10486. The choice of area makes a compromise between spatial resolution, range of depths and resolution in spatial wavenumber in the power spectra. A larger size will allow access to the deeper sub-surface layers, but only with coarser spatial resolution. On the other hand, a smaller size will not allow access to the deeper layers and also render the fitting of rings more difficult. The AR was located close to the disc center at the time of this flare, therefore, projection effects did not pose any serious difficulty in the analysis.

The 3-D Fourier transform of the data-cube gives a trumpet like structure in the frequency domain ($k_x,k_y,\omega$). A slice at fixed frequency $\omega$ of this structure gives power concentrated in concentric rings \citep{1988ApJ...333..996H}. These rings provide the characteristic properties of $p$-modes and subsurface flows. The phase velocity of the acoustic wave is augmented by flow velocity ($\textbf{U}$) causing change in frequencies ($\Delta \omega=\textbf{k.U}$) that perturbs the center and shape of the rings. Surface flows are derived from this perturbation using ring fitting. The $p$-modes of different wavelengths are trapped at different depths beneath the surface. Therefore, flows derived at the surface are weighted averages over depths. Inverting these characteristics of the modes give flows in the interior.

As illustrated in Figure \ref{fig:flare_datim}, we constructed five data-sets of 1664 minutes duration each with different starting time, such that the flare onset was placed in the beginning ($R_5$), one-fourth ($R_4$), center ($R_3$), three-fourth ($R_2$) and end ($R_1$) of the data-cube's time-line. The 16\arcdeg~patch consists of 128 pixels, giving a spatial resolution $\Delta x= 1.5184~\rm Mm$, i.e., the $k$-number resolution, $\Delta k=3.2328\times10^{-2} \Mmi$, and a Nyquist value for the harmonic degree, $l=1440$. The corresponding range in ($k_x,k_y$) space is from {-2.069\Mmi}~to 2.069\Mmi. The temporal cadence and duration of data-sets give Nyquist frequency of 8333\muHz~and frequency resolution of 10\muHz, respectively. 

To determine the surface mode parameters of \mbox{$p$-modes}, we have carried out ring-diagram analysis using the dense-pack technique \citep{2002ApJ...570..855H} adapted for GONG data \citep{2003ESASP.517..255C, 2003ESASP.517..295H}. Images were remapped around the central position (285\arcdeg, -22.5\arcdeg) using transverse cylindrical projection to obtain equi-distance spatial sampling interval required for Fourier transform. To remove the effect of differential rotation, the remapped images were tracked with differential rotation rate of the Sun \citep{1984SoPh...94...13S}. The tracked image cubes are apodized before being Fourier transformed. The main ring-fitted parameters (i.e., surface acoustic mode parameters) determined in our study are radial order ($n$), degree ($l$), mode amplitude ($A$), mode width (\mwd), zonal velocity (\Ux) and meridional velocity (\Uy). The common mode parameters in all sets were corrected for filling factor using a method described in \cite{2000ApJ...531.1094K}. Zonal (\ux) and meridional (\uy) components of sub-surface flows were derived by regularized least square (RLS) inversion from the surface to a depth of $\sim$20 Mm \citep{1996Sci...272.1300T, 2002ApJ...570..855H}.  

The topology of fluid is measured by kinetic helicity. It is defined as the volume integral of the dot product of the velocity ($\bf u$) and vorticity ($\bf\omega$) of the flows \citep{1992AnRFM..24..281M}:
\[ 
H_{\rm K}  = \int_V {{\bf u} \cdot {\bf\omega} \,dV} 
\] 
\noindent where, $\bf\omega  = \nabla  \times \bf u$ is the vorticity vector. It is a measure of circulations per unit area, i.e., twist of the flow, while its dot product with velocity gives variation of twist with flows. We estimated the vertical component of velocity $u_z$ from the divergence of horizontal components assuming mass conservation \citep{2005ApJ...630.1184K}. Now, having all three components of flows with depths, we can derive kinetic helicity. As we do not have the flows at each point of the active region from ring analysis, we can derive only the average helicity over the entire area, termed as helicity density ($h_K=\bf u \cdot\nabla\times\bf u$). The numerical derivatives of flows are derived using Lagrange three point interpolation. 

%
%
%
\section{Results and discussions}
\label{S-result}

\begin{figure}[ht] 
	\centering
		\includegraphics[width=0.95\textwidth,clip=,bb=30 5 554 407]{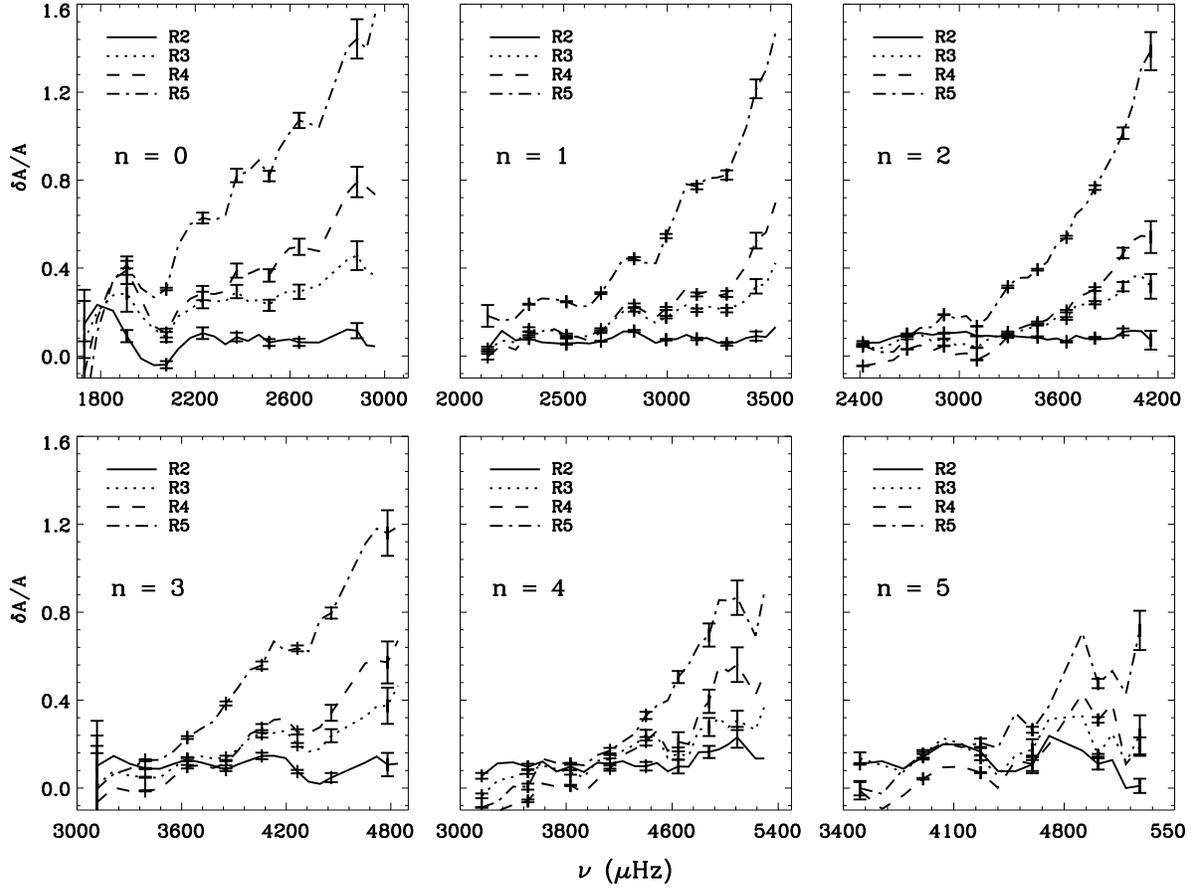}
		\caption{Relative difference in mode amplitude for  the data-sets  $R_i,  i=2, \ldots5,$ for radial orders, $n$ = 0 -- 5. Here, $R_1$ is used as the reference.}
	\label{fig:Amp}
\end{figure}

\begin{figure}[ht] 
	\centering
		\includegraphics[width=0.95\textwidth,clip=,bb=27 8 558 407]{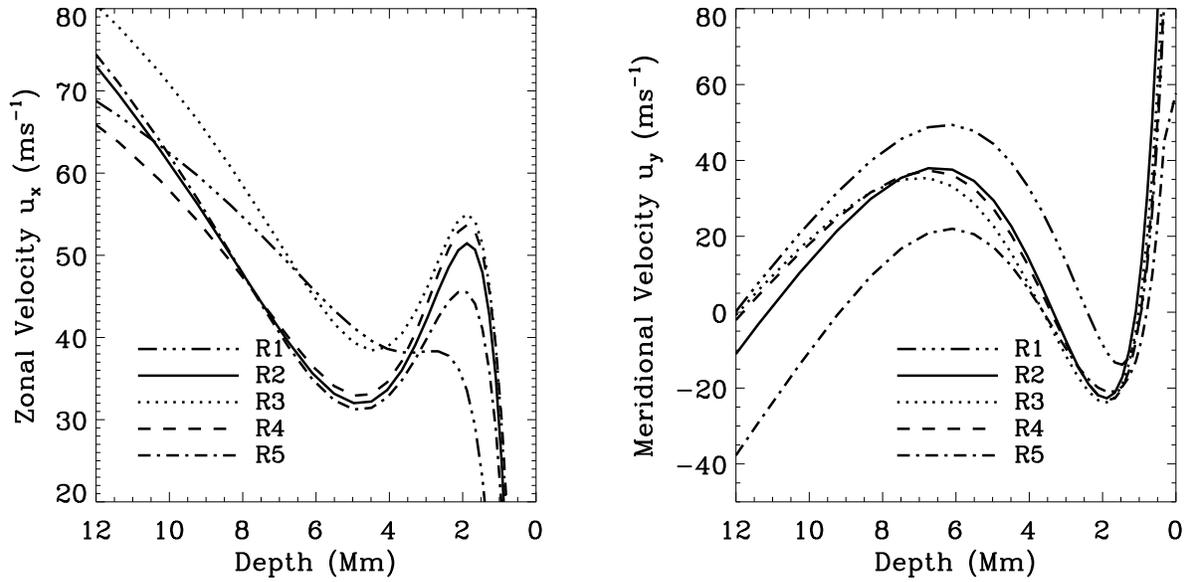}
		\caption{Zonal $u_x$ (left panel) and meridional $u_y$ (right panel) velocity profiles with depth from 0 to 12 Mm for the data-sets  $R_i,  i=1, \ldots5$.}
	\label{fig:invp}
\end{figure}

\begin{figure}[ht] 
	\centering
		\includegraphics[width=0.95\textwidth,clip=,bb=27 8 558  457]{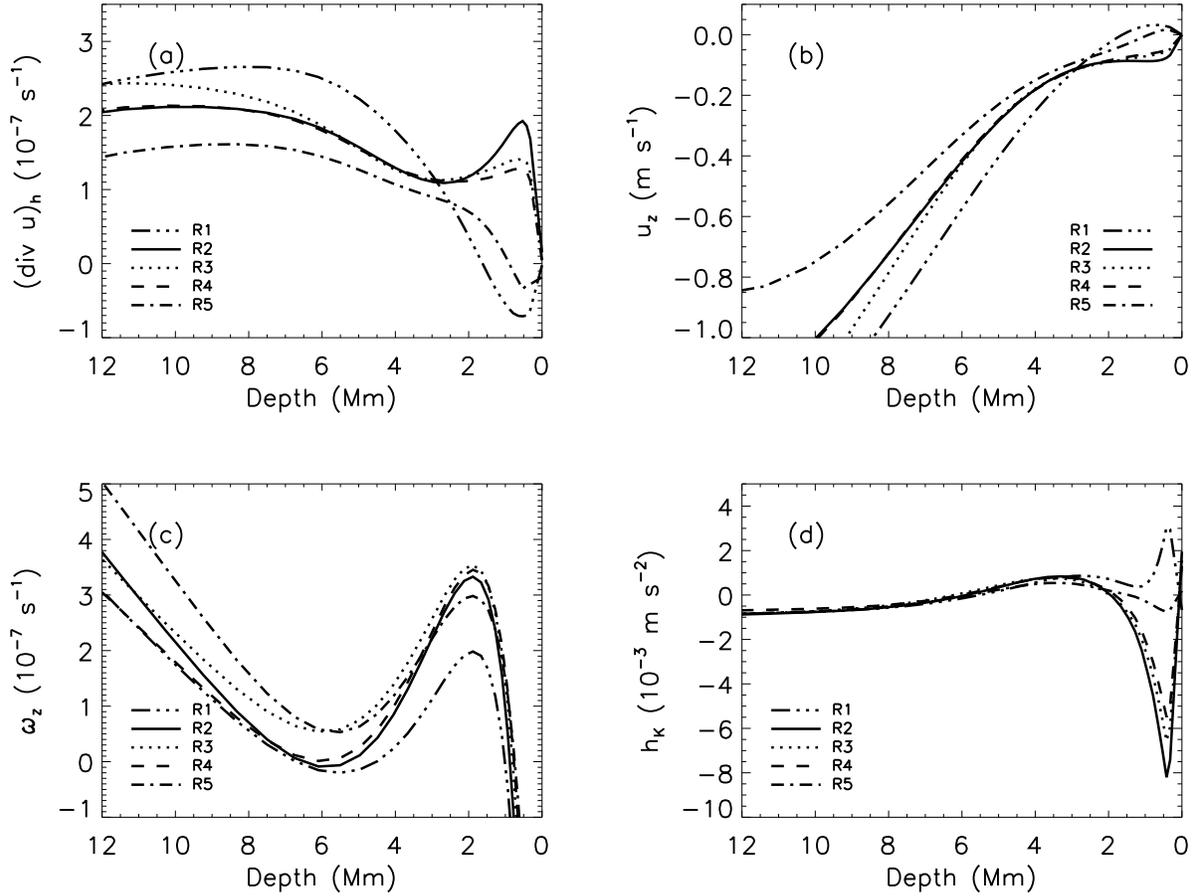}
		\caption{Inverted profiles of (a) divergence of horizontal components of velocity, (b) vertical component of velocity $u_z$, (c) vorticity and (d) helicity density, with depth from 0 to 12 Mm for the data-sets  $R_i,  i=1, \ldots5$.}
	\label{fig:invp2}
\end{figure}

Mode parameters obtained for the five data-sets are shown in Figure \ref{fig:flare_datim} for comparison of the flare-related effects. Surface acoustic mode parameters and the corresponding inverted parameters for these data-sets are shown in Figures~\ref{fig:Amp}--\ref{fig:invp2}. 

Figure~\ref{fig:Amp} shows that relative difference in mode amplitude obtained for $R_2$ to $R_5$ where $R_1$ is used as the reference. This quantity is found to be the largest for $R_5$ as compared to that for other sets. The difference between the amplitudes increased with frequency \mbox{$\nu > 2000~\mu$Hz}, and reduced with increasing radial order from $n$ = 0 to 5. The large difference of mode amplitude at lower radial orders may be attributed to the flare effect as these modes are confined near the surface and hence, more likely to be affected by the surface activity.  

The flare was placed at the beginning in $R_5$, hence the overall helioseismic contribution of this LDE event, extending from the pre- to post-flare phase, is expected to remain entirely within this data-cube. Correspondingly, the mode amplitude was found to be the largest in $R_5$ as expected. On the other hand, it was smallest in $R_1$ where the flare was placed at the end of the data-cube, therefore post-flare effects were not covered by this data-set. From the systematic trend of variations in the mode amplitude from $R_5$ to $R_1$, it is evident that the energetic flare indeed gave rise to a significant amplification of mode amplitude.
  
Mode width is also found to be large at all frequencies and radial orders for $R_5$. The changes are found to be larger at lower frequencies implying that higher-n modes have shorter life times.  \cite{1988IAUS..123...37D} and \cite{2009SoPh..258....1B} have also found that mode width increased with frequency as well as with degree of the modes. However, care is needed in this interpretation as significant contribution to mode width may arise due to the available limited resolutions in wavenumber and frequency. 

Using the ring diagram fits, we have calculated the horizontal velocities in order to check if there is any variation in surface velocities due to flares.  We found large variations in surface zonal and meridional velocities at all frequencies for the radial orders $n$ = 0 to 5 from $R_1$ to $R_5$. Also, the deviation for both the components of velocity  is larger for $R_5$ as compared to other data-sets indicating the variation in flows from the pre- to post-flare phases.

The fitted velocities for each mode can be inverted to calculate the zonal ($u_x$) and meridional ($u_y$) components of velocity as a function of depth. These were derived by regularized least square (RLS) inversion from the surface to a depth of 20 Mm \citep{1996Sci...272.1300T, 2002ApJ...570..855H}. Both $u_x$ and $u_y$ exhibited significant systematic changes with the flare. The profiles of zonal velocities $u_x$ with depth are shown for $R_1$ to $R_5$ in Figure~\ref{fig:invp} (\textit{left panel}). A large decrease in zonal velocity around the depth $\approx$ 4 Mm is common in all cases.  However the depth where the minimum of $u_x$ occurred varied from the shallowest for $R_1$ to the deepest for $R_5$. Figure~\ref{fig:invp} (\textit{right panel}) shows the coresponding changes in meridional velocities $u_y$ for $R_1$ to $R_5$. It is found that $u_y$ decreased rapidly from the surface to the depth of $\sim$ 1 Mm for all the data-sets. Thereafter, it increased and attained a peak at $\sim$ 6 Mm. It is inferred that the gradient in $u_y$ is largest for $R_1$ (i.e., before the flare) and it decreased systematically from $R_2$ to $R_5$ as the post-flare effects increased. This confirms the earlier reports about meridional velocity gradient by \cite{2004ESASP.559..293A}. 

Figure~\ref{fig:invp2} shows the profiles with depth of (a) divergence of horizontal components of velocity, (b) vertical component of velocity $u_z$, (c) vertical vorticity and (d) helicity density. Interestingly, it is observed that flare-related variations in helicity density $h_k$ are confined within a shallow region from the surface to a depth of around 3 Mm.  For depths greater than 3 Mm, $h_k$ derived for all the five cases approached toward the same value.

Changes in the topology of sub-surface flows during pre- to post-flare phases are illustrated in Figure~\ref{fig:invp2}a--d.  In $R_1$, the divergence of horizontal component ($u_h$) near a depth of \mbox{$\approx$ 1 Mm} changed from a small negative to a large positive value for $R_2$, and decreased from $R_3$ to $R_5$ (Figure~\ref{fig:invp2}a). This suggests that the perturbation in the flows at this depth gradually returned back to the  initial or pre-flare state after the flare decayed. On the other hand, the large positive divergence observed at depths below 3 Mm decreaseed in magnitude from $R_1$ to $R_5$. 

The vertical component of flows ($u_z$) changed from upward direction in $R_1$ to downward direction in $R_2$ at a depth of $\approx$ 1 Mm, and then back to the upward direction in $R_5$ through $R_3$ - $R_4$ (Figure~\ref{fig:invp2}b). The $u_z$ profiles converge to the same negative value ($\approx -0.1~\rm ms^{-1}$) at the depth of $\sim$3 Mm, and then diverges systematically from $R_1$ to $R_5$ as the depth increases. The vertical component of vorticity vector ($\omega_z$) is shown in Figure~\ref{fig:invp2}c for $R_1$ to $R_5$. The peak in $\omega_z$ near 2 Mm and the trough near 6 Mm manifest the bipolar structure of the AR. Kinetic helicity ($h_K$) changed from a small positive value for $R_1$ to a large negative value for $R_2$,  increasing toward the level of $R_1$ for $R_3$ -  $R_5$ (Figure~\ref{fig:invp2}d). However, it remained nearly constant for all the data-sets at depths below $\sim$3 Mm. These systematic changes in sub-surface parameters ($u_h$, $u_z$, $\omega_z$ and $h_K$) from $R_1$ to $R_5$ provide unambiguous evidence of the relationship of the large flare with sub-surface dynamics of the active region.

\section{Conclusions}
\label{S-conc}

This study of an extremely energetic and long duration X17.2/4B flare in NOAA 10486  gives a clear indication of flare-related variations in the $p$-mode parameters and sub-photospheric flows in NOAA 10486. Changing the temporal position of the flare within the Doppler data-cubes obtained for the active region amounts to the changing level of the pre- and post- effects in the data-set. The ring diagram analysis of the different data-sets thus constructed provides the following important results:

(i) The amplitude of $p$-modes increased up to $150\%$ in the case of the flare placed in the beginning of the data-set as compared to the case when flare was placed near the end or outside the data-cube. A similar result is obtained for $p$-mode energy also as expected due to its relation to the amplitude, manifesting the rate of energy supplied to the $p$-modes by the flare.

(ii) Amplitude and energy of the modes decreased with radial order indicating  that the effect of the flare decreased with increasing depth. Furthermore, we found that the amplitude and energy of modes increased with frequency. This suggests that modes with high natural frequencies are  amplified more by the flare as compared to the low frequency modes.

(iii) The gradient in meridional velocity observed at depths 2-6 Mm decreased with the flare which manifests relationship of the flare activity with the sub-surface flows.

(iv) The vertical component of the flow changed to the downward direction during the flare and then returned back to the pre-flare state. This is an evidence of downward moving material during the flare.

(v) Divergence of the horizontal component of flow near the surface changed from negative to positive and then back to the pre-flare state after the  flare. 

(vi) The sub-surface flow possessed a bipolar structure with one pole located at the depth of around 2 Mm and the other at 6 Mm. Near the surface, the flow was twisted during the pre-flare phase, which relaxed after the flare.

(vii) The kinetic helicity of sub-surface flow around the depth of 1 Mm changed from positive to negative during the pre- to the peak phases, and returned back to the pre-flare level after the post-flare phase.
 
In summary, this study provides strong evidence about the role of a large flare in modifying various acoustic mode parameters. Depending upon the temporal location of the large flare in NOAA 10486, systematic variations in mode parameters have been found for different data-sets with varying levels of pre- and post-flare effects. These results also suggest that flare-related changes in the acoustic mode parameters are detectable by ring diagram analysis provided that the pre- and post-flare phases of the flare (or flares) are well covered by the data-cube. Finally, we conclude that the sub-surface flow topology can be used as a proxy for flare-forecasting.

%
\acknowledgments

This work utilizes data obtained by the GONG program operated by AURA, Inc. and managed by the National Solar Observatory under a cooperative agreement with the National Science Foundation, U.S.A. The integrated X-ray flux data was obtained from GOES which is operated by National Oceanic and Atmospheric Administration, U.S.A. The authors would like to thank H.M. Antia for many useful discussions during the course of this work and providing mode inertia table for computing mode energy.



\newpage 


\end{document}